\documentclass[iop]{emulateapj}
\usepackage{graphicx}
\usepackage{amssymb,amsmath}
\usepackage{natbib}
\usepackage[backref,breaklinks,colorlinks,citecolor=blue]{hyperref}
\usepackage[backref,colorlinks,citecolor=blue]{hyperref}

\newcommand{\kA}{$\rm \kappa~And$~b}

\makeatletter

\newcommand{\Rmnum}[1]{\expandafter\@slowromancap\romannumeral #1@}
\makeatother

\shorttitle{The Water Abundance of $\rm \kappa~And$~b}
\shortauthors{Todorov et al.}
\begin{document}
\title{The Water Abundance of the Directly Imaged Substellar Companion $\rm \kappa~And$~\lowercase{b} Retrieved from a Near Infrared Spectrum}
\author{
Kamen O. Todorov\altaffilmark{1},
Michael R. Line\altaffilmark{2},
Jaime E. Pineda\altaffilmark{1,3},
Michael R. Meyer\altaffilmark{1},
Sascha P. Quanz\altaffilmark{1},
Sasha Hinkley\altaffilmark{4},
Jonathan J. Fortney\altaffilmark{2}
}
\altaffiltext{1}{Institute for Astronomy, ETH Z\"urich, Wolfgang-Pauli-Strasse 27, 8093 Z\"urich, Switzerland}
\altaffiltext{2}{Department of Astronomy \& Astrophysics, University of California, Santa Cruz, Santa Cruz, CA 95064, USA}
\altaffiltext{3}{Current address: Max Planck Institute for Extraterrestrial Physics, 85748 Garching, Germany}
\altaffiltext{4}{School of Physics, University of Exeter, Stocker Road, Exeter, EX4 4QL}

\begin{abstract}
  Spectral retrieval has proven to be a powerful tool for constraining the physical properties and 
  atmospheric compositions of extrasolar planet atmospheres from observed spectra, primarily for transiting 
  objects but also for directly imaged planets and brown dwarfs. Despite its strengths, this approach has 
  been applied to only about a dozen targets. Determining the abundances of the main carbon and oxygen-bearing 
  compounds in a planetary atmosphere can lead to the  C/O ratio of the object, which is crucial in understanding 
  its formation and migration history. We present a retrieval analysis on the published near-infrared spectrum of \kA, 
  a directly imaged substellar companion to a young B9 star. We fit the emission spectrum model utilizing a 
  Markov Chain Monte Carlo algorithm. We estimate the abundance of water vapor, and its uncertainty, in the atmosphere of the object. 
  In addition, we place an upper limit on the abundance of CH$_4$. We compare qualitatively our results to studies that have applied model retrieval on multiband 
  photometry and emission spectroscopy of hot Jupiters (extrasolar giant planets with orbital periods of several days) 
  and the directly imaged giant planet HR~8799b. 
\end{abstract}


\section{Introduction}
\label{sec:intro}

The study of exoplanets is moving from an era focussed on discovery 
to one of characterization. While most exoplanets detected to date 
have been discovered through the radial velocity or transit techniques, 
there are a handful of substellar companions within $\sim100$\,AU of 
their host stars discovered through direct imaging \citep{pep14}.
It remains unclear whether these populations are distinct or can 
be considered as part of a continuum of objects drawn from a 
variety of formation mechanisms, and experiencing a range of 
evolutionary histories. 

The compositions of transiting and directly imaged extrasolar giant 
planets may be an important marker of their
formation and migration history \citep{obe11,mad14a}. These authors 
suggest that giant planets' C/O and C/H ratios are affected by the 
locations in the protoplanetary disks where they formed with respect 
to the water and CO ice lines, as well as by the formation mechanism. 
E.g., \citet{obe11} suggest that stellar C/O and C/H ratios are indicative of
either formation within the water ice line or via gravitational instability. 
Substellar C/O, but superstellar C/H point to significant accretion of 
water ice bodies after the gas accretion phase. Superstellar C/O and C/H
can mean gas accumulation near the CO or CO$_2$ ice lines or significant
accretion of carbon grains. Superstellar C/O and substellar C/H indicates
gas accretion from outside the water snow line. A caveat is that \citet{obe11} assume that
the planet forms at a fixed location within the disk and do 
not consider the impact of disk migration.

The emission and absorption spectra of giant exoplanets and substellar companions 
can inform us about the chemical species and the C/O ratio in their atmospheres
and help resolve the debate about their formation mechanism and evolution. 
Two complementary approaches have been used in constraining the atmospheric 
structure and composition of these objects. The first one relies on sophisticated forward 
modeling of the physical and chemical processes in the atmospheres of giant planets and
calculating molecular abundances and pressure-temperature (P-T) profiles, and 
subsequently computing the resulting spectra \citep[e.g.,][]{for06, for08, bur07, bur08}. 
This approach, however, is computationally expensive,
which combined with the large number of free parameters makes it extremely difficult
to compare the model spectra to observations in a statistically robust manner. 
While model spectra that match the data can be found, quantifying the uncertainties 
in the underlying individual parameters is difficult. 

The other approach, ``spectral retrieval'', on which we focus here, 
relies on a simple radiative transfer model that assumes that the P-T profile,
the surface gravity and the atmospheric composition (with only few main molecular species 
considered) are free parameters \citep[e.g.,][]{ben12,ben13,lee12,lin12,lin13,hen16}. 
The effects of clouds on substellar emission spectra can also be 
parametrized and included \citep[e.g.,][]{mad11, lee13, lee14a, lee14b}. 
Model spectra produced in this way are computationally cheap 
and the limited number of parameters makes fitting to observations feasible. 
Since the physical phenomena included in these models are limited, they 
rely entirely on observational data to inform the fit, although priors based on 
theoretical considerations can improve the retrieval. 
A caveat is that the models used for retrieval, while still accounting for the most important 
physical processes and chemical species, could still be over-simplistic. Thus, not all 
features of a given atmosphere may be accounted for correctly. 

In this study, we examine an integral field near-infrared spectrum observed by \citet{hin13} of the directly imaged 
substellar companion orbiting the young star $\rm\kappa$ Andromedae \citep[hereafter \kA~following][]{car13}. 
We summarize the known properties of the $\rm\kappa$ And system in Section~\ref{sec:prop}.
In Section~\ref{sec:obs}, we discuss the available observed spectrum of \kA. 
We present our simple atmospheric model in Section~\ref{sec:mod}.  
Section~\ref{sec:ana} focuses on our spectral retrieval approach, 
and in Section~\ref{sec:dis} we discuss our results and compare them to similar studies
of giant hot exoplanets. 

\section{Known Properties of the $\rm\kappa$ Andromedae System}
\label{sec:prop}
$\rm\kappa$ Andromedae is a young but already post-main sequence B9IVn star \citep{wu11} whose
accurate age is debated. \citet{car13}, who discovered the companion, 
assume that $\rm\kappa$ And is a member of the Columba association \citep{zuc11} and,
following \citet{mar10}, adopt an age of $30^{+20}_{-10}$\,Myr for their analysis. 
On the other hand, \citet{hin13} use theoretical stellar isochrone tracks \citep{ber09} 
combined with the previously measured stellar properties to
derive an age of $220\pm100$\,Myr. An alternative analysis by 
\citet{bon14} adopts a more conservative value than \citet{car13}, but still 
based on the age of the Columba association: $30^{+120}_{-10}$\,Myr. 
The correct age of the system is important 
when assessing the luminosity and hence the mass of \kA, whose spectral type is $\rm L1\pm1$ \citep{hin13}. 
There is no available dynamical mass estimate for this object.
\citet{car13} use their estimate for the age of the system ($30^{+20}_{-10}$\,Myr) and the DUSTY
evolutionary models \citep{all11} to estimate a companion mass of $12.8^{+2.0}_{-1.0}M_J$.
On the other hand, \citet{hin13} adopt $\rm M_{comp} = 50^{+16}_{-13}M_J$, 
while \citet{bon14} place a lower limit on the 
mass of the object of $\rm M_{comp} > 10M_J$ based on ``warm-start'' evolutionary models. 
While in principle it is a basic parameter, the mass of the companion is 
of secondary importance for our purposes, because it enters the spectrum formation only via 
the surface gravity term, which also depends on the radius of the object. 
The effective temperature of \kA~is $2040\pm60$\,K \citep{hin13}. 
We summarize the known properties of the host star and its companion in Table~\ref{tab:prop}. 

   \begin{deluxetable*}{lll}
   \tabletypesize{\scriptsize}
\tablewidth{0pt}
\tablecaption{Stellar and Companion Parameters for the $\rm \kappa~And$ System}

\startdata
\hline\\[-1.5ex]
M$_\star$ (M$_{\odot}$) & $2.4 - 2.5$ & \citet{car13} \\
$\rm J_\star$ (mag)\tablenotemark{$\dagger$} & $4.624\pm0.264$ & 2MASS, \citet{skr06} \\
$\rm H_\star$ (mag)\tablenotemark{$\dagger$} & $4.595\pm0.218$ & 2MASS, \citet{skr06} \\
$\rm K_{s,\star}$ (mag)\tablenotemark{$\dagger$} & $4.571\pm0.354$ & 2MASS, \citet{skr06} \\
T$_{\rm eff,\star}$ (K)   &  $10730\pm250$ & \citet{wu11} \\
SpT$_\star$ & B9IVn & \citet{wu11} \\
Distance (pc) & $52\pm2$ &{\it Hipparcos}; \citet{per97};\\
		&		 & \citet{car13}\\
\hline\\[-1.5ex]
$\rm J_{comp}$ (mag) &  $16.3\pm0.3$ & \citet{car13} \\
$\rm H_{comp}$ (mag) &  $15.2\pm0.2$ & \citet{car13} \\
$\rm K_{s,comp}$ (mag)& $14.6\pm0.4$ & \citet{car13} \\
$\rm L^\prime_{comp}$ (mag)& $13.12\pm0.09$ & \citet{car13} \\
T$_{\rm eff, comp}$ (K) &  $2040\pm60$ & \citet{hin13} \\
\hline\\[-1.5ex]
Projected separation (\arcsec)& $1.064\pm0.006$ & \citet{car13}\\
Projected separation (AU)& $56\pm2$ & \citet{car13}\\
Position angle ($^\circ$) & $55.9\pm0.4$ & \citet{car13}
\enddata
\tablenotetext{$\dagger$}{Two Micron All Sky Survey (2MASS) magnitude of the star (from the Infrared Science Archive: 
\url{http://irsa.ipac.caltech.edu}).}
\label{tab:prop}
\end{deluxetable*}

\section{Observations}
\label{sec:obs}
In our analysis, we consider the spectrum of \kA~as published by \citet{hin13}. 
We offer a brief summary of how the data were obtained.  
These authors used the ``Project 1640'' instrument \citep{hin11, opp12} on the
200\,inch Hale Telescope at Palomar Observatory. Project 1640 is an integrated
combination of a coronagraph and an integral field spectrograph (IFS) and 
covers the Y, J and H bands. The data were obtained on UT 2012 December 23, 
starting at an airmass of 1.02. A total of 16 images were taken with exposures of
183\,s each, with the Hale Telescope adaptive optics on, and the star hidden 
behind the coronagraphic mask. The spectrum extraction is described in detail 
in \citet{hin13}. The observations cover the range between $0.9$ and $1.8\,\mu$m,
but for this study, we focus on wavelengths longer than $1\,\mu$m. At shorter 
wavelengths, the molecules that we consider (CH$_4$, CO, CO$_2$, H$_2$O,
C$_{2}$H$_{2}$, HCN, NH$_{3}$, see Section~\ref{sec:mod}) and
the P-T profile have a limited impact on the emission spectrum \citep[e.g.,][]{lee14a}.
The spectrum is low-resolution ($\rm R\approx45$) and contains 28 flux points in the
range we examine. 
The uncertainties derived by the observers for every spectroscopic channel 
include the uncertainties due to photon noise, systematic errors and uncertainties resulting from the spectral calibration.
Unfortunately, the absolute calibration of the resulting flux values is not precise, and therefore it is difficult to derive
the radius of the companion from its luminosity using the well known distance to $\rm\kappa$~And 
\citep[$52\pm2$\,pc from {\it Hipparcos} parallaxes;][]{per97,car13}.

\section{Atmospheric Emission Model}
\label{sec:mod}
\subsection{General Description}
\label{sec:gendesc}
For this study we require a relatively fast and therefore relatively simple radiative transfer code
that takes a limited number of parameters -- chemical species volume mixing ratios, P-T profile and a
scaling factor that incorporates the companion's distance and luminosity -- and computes an emission spectrum. To this end, 
we implement a custom one-dimensional plane-parallel atmospheric model that 
we validate against the results from the NEMESIS code \citep{irw08}
and the spectral synthesis software used by \citet{lin12} and \citet{lin13}. 
These codes were designed to study the spectra of planets, including hot Jupiters and 
young directly imaged gas giants. While we occasionally refer to them as ``radiative transfer'' models for simplicity, we 
do not enforce radiative equilibrium in our code, since \kA~is a self-luminous object releasing its 
primordial heat of accretion. Due to its separation from the primary, the companion is negligibly externally heated. 

Following \citet{lin12}, we consider the effects of nine chemical species on the emission spectrum in our 
radiative transfer model: CH$_4$, CO, CO$_2$, H$_2$O, H$_2$ and He, as well as 
C$_{2}$H$_{2}$, HCN and NH$_{3}$. The first four are expected to be
the major sources of molecular line opacity in a hot substellar atmosphere \citep[e.g.,][]{tin07, swa09}.
Other relatively common molecules also have absorption bands in the wavelength range we consider, e.g.
hydrogen cyanide (HCN),  acetylene ($\rm C_{2}H_{2}$) and ammonia \citep[$\rm NH_{3}$; e.g.,][]{rot09}. 
The abundances of these species in self-luminous gas giant planets and brown dwarfs are 
expected to be much lower compared to the abundances of water, carbon monoxide, carbon dioxide 
and methane \citep[e.g.,][]{lod02, zah14}, but nevertheless could have an impact on the emitted spectrum and therefore we explore models that
include these three species. Alkali metal atomic species, e.g., K I and Na I, have some lines that fall within the wavelength range we consider.
To investigate this, we compare the \kA observation with an observation from the SpeX Prism Spectral 
Libraries\footnote{\url{http://pono.ucsd.edu/~adam/browndwarfs/spexprism/}} of a 
L2.5 brown dwarf with similar effective temperature \citep[2MASS J14343616+2202463;][]{she09}. 
In a typical example, it is possible that the \kA~ observation at $1.14\,\mu$m is 
affected by a Na I line doublet. However, we would need signal to noise 3-5 times higher to be able to make this claim. 
Thus, due to the limited resolution and large uncertainties of our data, neutral atomic alkali lines are unlikely to have a significant impact on our results.  

For CO, CO$_2$, H$_2$O we adopt opacities from the HITEMP database \citep{rot10}. 
For methane, acetylene, hydrogen cyanide and ammonia, we adopt the opacities from \citet{fre14}.
In addition, we include the collision induced opacity due to H$_2$-H$_2$ and H$_2$-He interactions \citep{bor01, bor02}. 
The adopted opacities cover the temperature range between 500 and 3000\,K and the pressure range between $10^{-6}$ and $10^{2}\,$bar.
The line-by-line cross section databases as a function of T and P are sampled at a resolution of $\rm 1\,cm^{-1}$.
Throughout this study, we consider the abundance of each species as a fraction of the total number of molecules. 
The CH$_4$, CO, CO$_2$ and H$_2$O abundances are free parameters in all of our fits. 
We run fits with the abundances of C$_{2}$H$_{2}$, HCN, NH$_{3}$ all fixed at zero or all set to be free.
The ratio between H$_2$ and He is fixed to 86:14 \citep{lin12}. 
The mean molecular weight of the atmosphere is calculated based on these main 
chemical species, assuming that the contribution from other molecules is small. 

In our model, the atmosphere is divided into a number of horizontal layers as a function of
pressure. In this study, we use 90 layers that are evenly spaced in $log(P)$, 
varying between $2.4\times10^{-6}$ and $\rm 163.3\,$bar. 
Each layer has a temperature, based on the
input P-T profile and a chemical composition associated with it. Following \citet{lin12}, 
we keep the relative abundances by number of all molecules we consider uniform for all layers, 
since the information content of the spectrum is 
insufficient to allow the retrieval of chemical gradients. We calculate the amount of flux from a given layer 
that leaves the atmosphere unimpeded and integrate over all layers, for a given wavelength. 
In other words, like \citet{lin12}, we follow, e.g., \citet{gra05} and compute, for each atmospheric layer, the 
change in optical depth, 
\begin{equation}
	\Delta \tau_{i,\,\lambda} =  \frac{\Delta P_{i}}{\mu\,g} \sum_{k=0}^{N}(\sigma_{\lambda,\,k}\,f_{\lambda,\,k}),
\label{eqn:tau}
\end{equation}
where $\Delta P_{i}$ is the change in pressure in the $\rm i^{th}$ layer (in units of barye, i.e., $\rm g\,cm^{-1}\,s^{-2}$); 
$\mu$ (in $\rm g\,mol^{-1}$) and g (in $\rm cm\,s^{2}$) are the mean molecular weight 
and the surface gravity, respectively. $\sigma_{\lambda,\,k}$ (in $\rm cm^{2}$) and $f_{\lambda,\,k}$ are the molecular cross-section
 at wavelength $\lambda$ and mixing ratio of the $\rm k^{th}$ chemical species, respectively. We integrate this value over all layers
lying above the $\rm i^{th}$ layer to calculate the total amount of optical depth, $\rm \tau_{i,\,\lambda}$, that a photon emitted from this layer would encounter.
Since some of the light emitted by any given layer will be absorbed within that layer, we account for this effect by dividing the $\rm i^{th}$
layer in upper and lower half in pressure by dividing $\rm \Delta P_{i}$ by two and adding
the $\Delta \tau_{i,\,\lambda}$ value of the upper half to $\rm \tau_{i,\,\lambda}$. The emitted intensity of a ray of light 
through the atmosphere is then, 
\begin{equation}
	I_{\lambda} = \sum_{i=1}^{90} B(T_{i}, \lambda) e^{-\tau_{i,\,\lambda} / \cos(\theta)} \frac{\tau_{i,\,\lambda}}{\cos(\theta)}, 
\label{eqn:int}
\end{equation}
where B is the Planck black body radiation at temperature $\rm T_{i}$ and $\theta$ is the angle between the direction to the
observer and the normal to the surface. In order to get the observable flux, we integrate this over the visible surface of the planet, 
\begin{equation}
	F_{\lambda} = \int_{0}^{2\pi}\int_{0}^{\pi}I_{\lambda}\cos(\theta)d\theta d\phi, 
\label{eqn:bb}
\end{equation}
To compute this integral efficiently, we take advantage of the Gaussian quadrature approximation.
The same operation is performed for every wavelength in which we are interested between 
$1$ and $1.8\,\mu$m. 
The layers at the top of the atmosphere have too low optical depths to have a significant impact on the spectrum 
in the near-infrared, while the layers near the bottom are too obscured by the opacity of 
the upper layers (the top 10 and bottom 10 layers have $<0.001\%$ contribution to the
total flux). 

We sample the model spectrum at a resolution 50 times higher than that of the observed spectrum -- 
every $\rm \sim 3\,cm^{-1}$, or $\rm \sim 6\,\AA$ in this wavelength range. The model emitted flux is 
binned in wavelength to match the resolution of the observed spectrum. 
This approach is similar to the methods of \citet{lin12} and \citet{lin13}.  
Since the molecular opacity data is 
sampled at $\rm 1\,cm^{-1}$, finer sampling than that will not improve the information content of our model. Sampling 
our model spectrum at full $\rm 1\,cm^{-1}$ resolution slows down the model calculation by a factor of three compared to 
sampling at our choice of $\rm 3\,cm^{-1}$ and makes Bayesian fitting impractical. To ensure robustness, we produce test model spectra sampled 
at $\rm 1\,cm^{-1}$ and $\rm 3\,cm^{-1}$ for several arbitrary but reasonable parameter combinations and find that,
after binning the models to the resolution of the data, they are different by $\lesssim1\%$ per spectroscopic point, 
much less than the uncertainties of the observed spectrum. Therefore, our choice of wavelength sampling of the models
does not impact our results. 

\subsection{Treatment of Clouds}
\label{sec:clouds}

\begin{figure}
  \centering
  \includegraphics[scale=0.3]{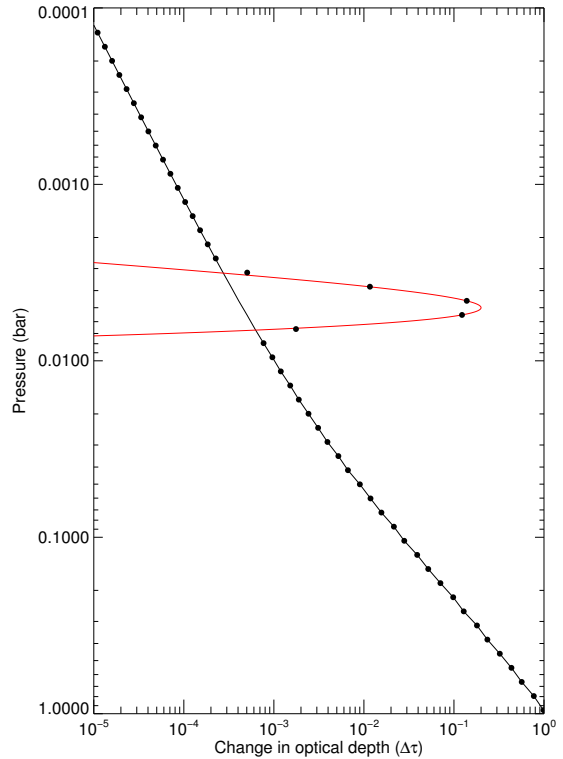}
  \caption{We show a {\it schematic} representation of our cloud treatment approach for an arbitrary 
    combination of molecular mixing ratios and at an arbitrary wavelength. The relation between pressure and 
    optical depth is represented by the black line. 
    The red line is a Gaussian describing the contribution to the change in optical depth due to the 
    cloud layer. The pressure at the peak of the Gaussian, its amplitude and standard deviation
    are free parameters in the fit. The black points show the total change in optical depth in a given atmospheric layer.}
  \label{fig:clouds}
\end{figure}

Previous emission spectrum retrieval studies have treated clouds and hazes \citep{lee13,lee14a,lee14b}. 
These authors have explored cloudless and uniformly cloudy (the atmosphere being uniformly permeated 
by particles of a given size) models as well as an ``intermediate'' model, where the cloud layer has 
a minimum and maximum pressure, similarly to \citet{bur11}. 
\citet{lee14a} have discussed aerosol treatment in detail and included it in their
retrieval for the transit spectrum of HD~189733b by varying the wavelength dependent
number density of the aerosol particles and their size.  We adopt a more numerically simplified approach, choosing
to parametrize the macroscopic quantity of cloud optical depth instead of the microscopic quantities 
of cloud particle opacity and number density.
Since our radiative transfer code is one-dimensional, we consider a single cloud 
layer that covers the whole planet with no gaps. 
The clouds are represented by adding ``grey opacity'' to the molecular opacities
in the atmospheric layers in the model where the cloud occurs.
This is usually equivalent to assuming that the cloud particles are much larger than the observation wavelength, 
but since we consider a relatively narrow wavelength range between 1 and 1.8$\,\mu$m, there are cloud chemical species 
that would appear approximately grey even if their particles are comparable in size to the observed wavelength \citep[e.g.,][]{mor12,lee13}.
The amount of radiation that passes through the cloud is determined by the amount of extra optical depth contained in the ``cloudy'' atmospheric layers. 
The optical depth contribution from the cloud deck is invariant with wavelength.
We describe the optical depth contribution of the cloud deck as a Gaussian as a 
function of pressure in log space, similarly to \citet{hen12}. The altitude of the deck (i.e., the pressure 
at the location of the peak of the Gaussian), the peak optical depth contribution of the cloud and the 
vertical extent of the cloud are emission model inputs and can be explored by the 
fitting routine as free parameters. This scheme is illustrated in Figure~\ref{fig:clouds}. 
It is possible that the cloud permeates a ``wide'' range of pressures, i.e., 
the standard deviation of the Gaussian optical depth contribution is so large that the Gaussian 
is in essence flat in our region of interest. This case would be equivalent to the uniformly 
cloudy model in \citet{lee13, lee14b}. 

For simplicity, we do not include the effect of scattering that will effectively reduce the ``apparent" opacity. 
Forward scattering particles will transport more light through the top of the cloud, which is the equivalent of 
reducing the cloud opacity in pure absorption.  Thus, scattering and cloud opacity will be degenerate in our 
simple framework. For a detailed theoretical discussion of clouds in brown dwarfs and young planets, see \citet{mor12}.

\section{Fitting Procedure}
\label{sec:ana}

\subsection{Free Parameters}
\label{sec:freepar}
We explore four cases of the model fit to the observed spectrum -- 1) without clouds and only including the four main 
molecules: CH$_4$, CO, CO$_2$ and H$_2$O; 2) cloudy and only including the four main 
molecules; 3) cloud-free, including the four main molecules plus the three additional ones: 
C$_{2}$H$_{2}$, HCN, NH$_{3}$; and 4) cloudy and all seven considered molecules. 
We summarize the parameters and their definition in 
Table~\ref{tab:param}. For each case, we indicate which parameters are left free and which are fixed at 0. 
The spectrum model requires that 90 layers in pressure and temperature are 
specified. However, the temperature cannot be let 
to vary freely in each layer, because the spectrum does not contain sufficient information for this -- even neglecting the
signal-to-noise considerations, there would be more than 90 free parameters compared to a spectrum of 28 points. 
We select nine individual layers, uniformly interspersed in pressure, where the temperatures are free parameters ($\rm T_1-T_9$).
We then use quadratic interpolation to determine the temperature values for the 81 remaining layers between these points. 
In order to test whether a 9-point P-T profile is reasonable, we run two MCMC chains (one million iterations each) with clouds turned off and by fixing $log(g_{surf})=4.5$
using a 7-point and a 12-point P-T profile. We find that the molecular volume mixing ratio values resulting from these are consistent with the MCMCs using a 
9-point model well within $1\sigma$. 

The surface gravity of the companion, $g_{surf}$, is uncertain, and expected to be correlated with the molecular abundances. 
High surface gravity flattens the spectrum and requires a higher abundance of a molecular species to produce a feature of a given depth. 
Thus, fixing the surface gravity to a ``reasonable'' value would result in chemical abundance estimates with inaccurate uncertainties.
Hence, even though we cannot constrain this property, we leave it as a free parameter in all four cases that we explore in detail, 
such that $\rm log(g_{surf}) \in (3.5, 5.2)$, in cgs. We pick this interval based on estimates of the surface gravity of \kA~from \citet{hin13}.

While the relative flux from point to point within the spectrum is well calibrated, the absolute flux calibration is uncertain. 
To compensate for this and for the unknown radius of \kA, we introduce a multiplicative term, $\rm A_{calib} \sim \frac{R_{comp}^2}{d^2} G_{conv}$, 
where $\rm G_{conv}$ is the unknown conversion factor between measured and ``true'' flux. $\rm R_{comp}$ is the radius of the companion and d is
the distance to the $\kappa$ And system.

   \begin{deluxetable*}{lccccl}
   \tabletypesize{\scriptsize}
\tablewidth{0pt}
\tablecaption{Parameters of the Fit}
  \tablehead{
    \colhead{Parameter} &
    \colhead{Case 1\tablenotemark{$a$}} & 
    \colhead{Case 2\tablenotemark{$b$}} & 
    \colhead{Case 3\tablenotemark{$c$}} & 
    \colhead{Case 4\tablenotemark{$d$}} &  
    \colhead{Description} \\
    \colhead{(units)} &
    \colhead{}&
    \colhead{}&
    \colhead{}&
    \colhead{}&
    \colhead{} } 
\startdata
{\it Main molecules }\\
\hline\\ [-1.0ex]
$\rm n_{H_2O}$ & free & free & free & free & Water molecules number fraction. \\
$\rm n_{CH_4}$ & free & free & free & free & Methane molecules number fraction. \\
$\rm n_{CO} $ & free & free & free & free & CO molecules number fraction. \\
$\rm n_{CO_2} $ & free & free & free & free & $\rm CO_2$ molecules number fraction. \\
\\
{\it Other molecules } \\
\hline\\ [-1.0ex]
$\rm n_{C_{2}H_{2}}$ & 0 & 0 & free & free & $\rm C_{2}H_{2}$ molecules number fraction. \\
$\rm n_{HCN} $ & 0 & 0 & free & free & HCN molecules number fraction. \\
$\rm n_{NH_{3}} $ & 0 & 0 & free & free & $\rm NH_{3}$ molecules number fraction. \\
\\
{\it Cloud parameters} \\
\hline\\ [-1.0ex]
$\rm C_{amp}$  & 0 & free & 0 & free & Cloud peak optical depth; amplitude in Fig.~\ref{fig:clouds}.\\
$\rm C_{press}$ & \nodata & free & \nodata & free & Pressure where the cloud layer is centered. \\
$\rm C_{vert}$   & \nodata &  free & \nodata & free & Vertical extent of the clouds (bar). \\
\\
{\it Physical properties} \\
\hline\\ [-1.0ex]
$\rm g_{surf}$ ($\rm cm\,s^{-2}$) & free & free & free & free & Surface gravity, $\rm log(g_{surf}) \in (3.5,5.2)$. \\
$\rm T_1-T_9$ ($\rm K$) & free & free & free & free & Temperatures throughout the atmosphere. \\
\\
{\it Calibration} \\
\hline\\ [-1.0ex]
$\rm A_{calib}$ & free & free   & free & free & Multiplicative offset (poor absolute calibration).
\enddata
\tablenotetext{$a$}{Cloud-free, including only the four main molecules.}
\tablenotetext{$b$}{Cloudy, including only the four main molecules.}
\tablenotetext{$c$}{Cloud-free, including all seven molecules.}
\tablenotetext{$d$}{Cloudy, including all seven molecules.}
\label{tab:param}
\end{deluxetable*}

\subsection{Markov Chain Monte Carlo}
As a first step in the fitting process, we employ the IDL MPFIT $\rm \chi^2$-minimization library \citep{mar09} 
to provide us with an initial guess about the location of the minimum $\rm \chi^2$ in 
the free parameter space.  
Then, we employ a Markov Chain Monte Carlo (MCMC) approach to get best fit values and uncertainties. 
For this analysis, we have utilized the MCMC algorithm implemented and described by \citet{tod12, tod14}, who 
follow \citet{ford05,ford06}. We use the Metropolis-Hastings algorithm within
the Gibbs sampler in conjunction with a Gaussian trial value probability distribution.
We assume flat priors for all parameters.
During a given MCMC iteration, we perturb only a single randomly 
chosen free parameter. Each MCMC chain that we execute has $10^6$ iterations, or approximately
50,000-70,000 iterations per parameter. The initial part of the MCMC chain (10\% or $10^5$ iterations)
is discarded as ``burn-in'' time necessary for the algorithm to converge. Before running a long chain, 
we run shorter chains in order to choose the ``step sizes'', i.e., the standard deviations of 
the Gaussian probability distributions used to select the amount by which a given parameter is 
perturbed. In order to optimize the convergence speed of the chain, we select, by experimentation, 
the Gaussian widths such that the acceptance rate of the given parameter is between 15 and 35\%.
We constrain the molecular number fractions of the molecules to be less than $10^{-1}$ and the 
temperatures are constrained to be less than 3000\,K by preventing any trial states that exceed
these values. This is done in order to prevent the chains from entering temperature regimes where
the opacities are not defined and preventing the total abundance of the molecules included in the fit
from becoming unrealistically large. 

\section{Results and Discussion}
\label{sec:dis}
\subsection{Adopted Parameter Values and Uncertainty Estimation}
\label{sec:unc}

\begin{figure*}
  \centering
  \includegraphics[scale=0.5]{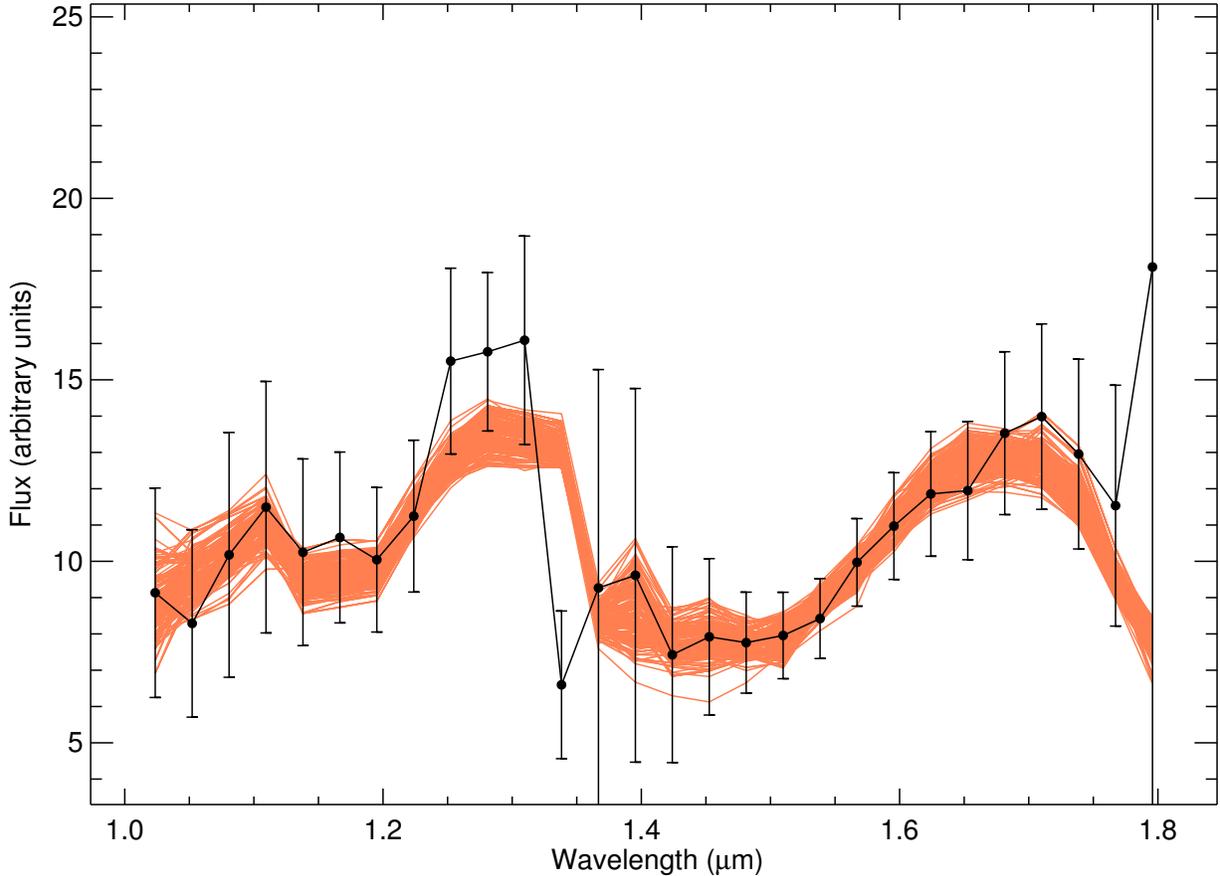}
  \caption{ We compare about 300 instances of the model cloud-free four-core-molecule MCMC run (orange lines) to 
    the observed spectrum of \kA~\citep[connected black points;][]{hin13}. Each shown model 
    is randomly chosen from a set with $\Delta\chi^2 < 1$ compared to the model with 
    minimum $\chi^2=16.1$ ($\rm \chi^2_{reduced}=1.25$). The water absorption 
    feature at $1.4\,\mu$m is clearly seen in the spectrum. While we plot absolute units on the vertical axis, 
    the absolute calibration of the spectrum was poor in the original observations (unlike the relative calibration
    at different wavelengths). Thus, the ordinate axis here should be used to compare 
    the difference of the fluxes at different bands and not the absolute flux of \kA. The details of the observed 
    spectrum and its uncertainties are discussed in Section~\ref{sec:obs}.
  }
  \label{fig:bestfit}
\end{figure*}

As discussed in Section~\ref{sec:freepar}, we run an MCMC chain for each of the four cases
we consider -- cloudy vs. cloud-free and four main molecules vs. all seven species in the model atmosphere.  
For all chains, the best fit model is chosen as the model with the minimum $\chi^2$ value in the MCMC chain, after
the initial $10^5$ iterations have been dropped as mentioned before. While this 
model best fits the available data, its parameters are not necessarily the values on which 
the MCMC converges finally. 
We find that the reduced $\chi^2$values for the fits are 1.25, 2.21, 1.13 and 2.06 for cases 1-4 
(as defined in Table~\ref{tab:param}), respectively. The corresponding Bayesian information criterion 
(BIC) values are 66.1, 82.1, 71.3 and 84.4. 
Even though the reduced $\chi^2$ values of the best fit models
that the four-species chains encountered are larger than the values produced by the seven-species fits, 
the BIC values for the latter models 
are significantly larger. Similarly, cloud-free BIC values are much smaller than cloudy ones. 
This suggests that the improvements in the fit achieved by adding molecules and clouds are not significant given the extra free parameters 
in the model, and for our results discussion, we adopt the outcome of the chain with the minimum BIC,  
case 1: cloud-free, including only CH$_4$, CO, CO$_2$ and H$_2$O.
We compare the best-fitting emission models from this MCMC run
to the data in Figure~\ref{fig:bestfit}. 

The surface gravity does not converge in any of the tested chains and the MCMC histograms for this 
parameter are flat. Therefore, we are unable to place any constraints on it.

We present the P-T profiles for case 1 in Figure~\ref{fig:bestfitPT}, and for all cases in the Appendix, Figure~\ref{fig:bestfitPTall}.
Several percent of the P-T profiles tested, especially for case 3, cloud-free with seven molecules, exhibit sharp temperature inversions. 
We find that the affected model atmospheres contain very little water 
($\rm log(n_{H_2O})\lesssim-6$) but the abundances of other molecules tend to be large (methane, CO and CO$_{2}$ have 
$\rm log(n)\gtrsim-2.5$)). The temperature inversions cause these molecules to 
appear in emission instead of absorption. Since there is no water in these models, there is a trough (in emission) near 1.4um, while on either 
side the CO, CO$_{2}$ and CH$_{4}$ can be seen in emission. The overall flux level of the spectrum is then compensated by the $\rm A_{calib}$ free parameter
(Table~\ref{tab:param}). 
At the resolution of the observations, this configuration closely resembles the observed spectrum.  
The inverted P-T profiles represent likely unphysical states of our vector of free parameters, considering the  strength of the required temperature inversion, 
that nevertheless result in spectra similar to the observed. This underscores the importance of high-resolution spectra of substellar companions for accurate 
atmospheric composition determination. These states are permitted in our simplified 
model fitting framework that does not impose any restrictions from atmospheric heat transport or chemical equilibrium. 
Were they the majority or a large fraction of all MCMC states, especially in case 1, 
we would conclude that the observed spectrum does not contain sufficient information to constrain the atmospheric properties 
using our approach. However, they constitute a minority of the MCMC states. 
Thus, we conclude that the inverted P-T profile states have little influence on our final results.

\begin{figure}
  \centering
  \includegraphics[scale=0.3]{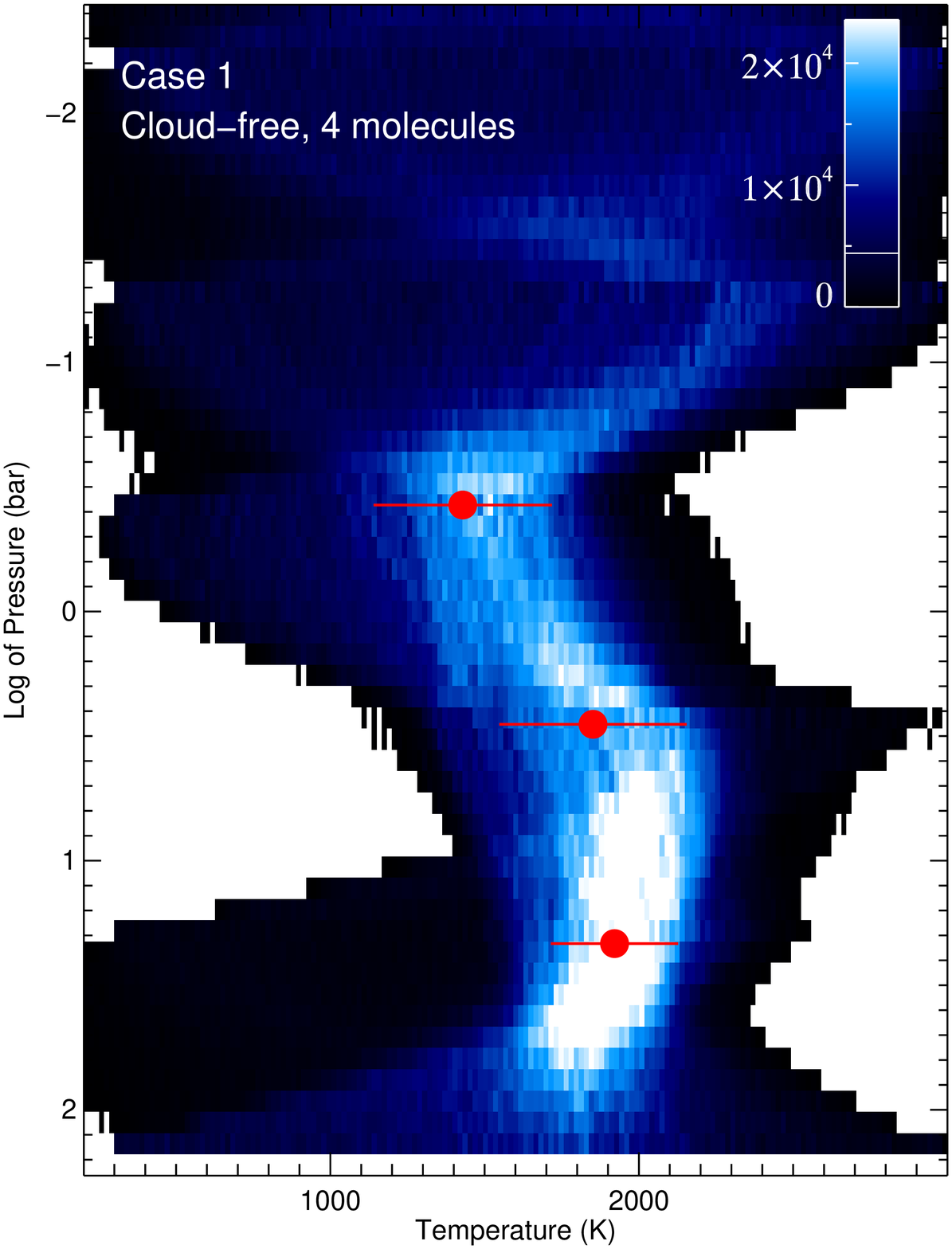}
  \caption{ Histogram of the temperature values at each pressure level from the MCMC run
    for case 1: without clouds and only including CH$_4$, CO, CO$_2$ and H$_2$O.
    The best-fit temperatures at pressures corresponding to the $\rm T_{1}-T_{9}$ free parameters
    are shown as red points with error bars. The uncertainties are based on the corresponding MCMC histograms. We define 
    temperature parameters with uncertainties larger than 500\,K as unconstrained (not shown). 
  }
  \label{fig:bestfitPT}
\end{figure}

We estimate uncertainties and adopt best fit values based on the histograms of the MCMC
parameter states. 
Even though we adopt the case 1 converged parameter values for our discussion based on the 
BIC values for the best fit models from the four cases, we present the results for both the cloudy and cloud-free models
in the Appendix (Table~\ref{tab:resultsall} and Figures~\ref{fig:bestfitPTall}, \ref{fig:h2ohistall} and  \ref{fig:ch4histall}). 

\begin{figure}
  \centering
  \includegraphics[scale=0.6]{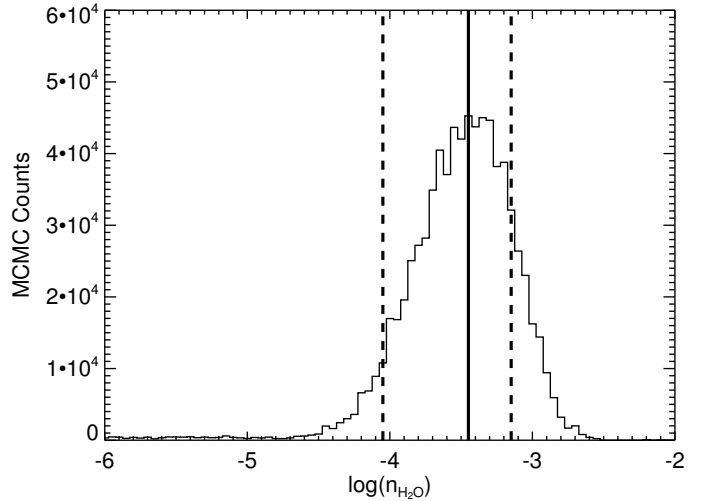}
  \caption{The histogram of cloud-free MCMC parameter parameter states for the molecular number fraction of water in log space
    for the cloud-free case, including only the four main molecules(case 1). The MCMC histogram peaks strongly near $\rm log(n_{H_2O}) = -3.5$. 
    We denote the peak of the distribution for case 1 (solid line) and the region that covers 34\% (``1$\,\sigma$'') of the values on either
    side (dashed lines). 
  }
  \label{fig:h2ohist}
\end{figure}

We are able to estimate the volume mixing ratio of water molecules
and we show the cloud-free water abundance histograms in Figure~\ref{fig:h2ohist} for case 1, 
although the results are consistent between cases (Table~\ref{tab:resultsall}). 
The case 1 histogram for CH$_4$ is flat, indicating poor convergence, but has a 
sharp peak and then cut-off towards higher number fractions. The cut-offs is well below 
the maximum value of $0.1$ permitted in the MCMC. 
Thus, we place upper limits on the abundance of CH$_4$ by 
reporting the number density value below which 95\% of MCMC iterations have occurred 
for that parameter. The abundances of CO and CO$_{2}$ remain unconstrained. 
The water volume mixing ratio and the upper limit for methane's volume mixing ratio
are consistent with predictions by \citet{hub07} for brown dwarfs, and with estimates by
 \citet{hen16} for C/O=0.5 at $\rm P=1\,$bar, assuming solar metallicity.

Both CO and CO$_{2}$ have absorption bands in the 1 to 1.8$\,\mu$m range, 
but these bands are weak, compared to water at these temperatures and pressures \citep[HITEMP database][]{rot10}, 
therefore, it is not surprising that these species, along with C$_{2}$H$_{2}$, HCN, NH$_{3}$ have little impact on our results, 
and especially the measured water volume mixing ratio. 
For all cloudy and cloud-free models, the water fractions are consistent with each other
suggesting that clouds do not dominate the emission spectrum of the companion at this wavelength range, 
in the context of our simple cloud model. 

The P-T profiles we present in Figure~\ref{fig:bestfitPT} do not encompass all of the nine 
levels in pressure where the temperature is a free parameter. The temperatures
at altitudes above the $\rm \sim 0.1\,bar$ level do not converge, since these layers are physically 
located well above the part of the atmosphere where the optical depth is close to 1, for all wavelengths 
between 1 and 1.8\,$\mu$m.

   \begin{deluxetable}{lcccc}
   \tabletypesize{\scriptsize}
\tablewidth{0pt}
\tablecaption{Results from the Case 1 MCMC Run}
  \tablehead{
    \colhead{Parameter\tablenotemark{a}} &
    \colhead{Best Value\tablenotemark{b}} 
    	} 
\startdata
$\rm log(n_{H_2O})$                                           & $-3.5^{+0.3}_{-0.6}$ \\ 
$\rm log(n_{CH_4})$\tablenotemark{c}                & $<-2.6$ \\
$\rm log(n_{CO}) $                                              & no constraint \\ 
$\rm log(n_{CO_2})$                                            & no constraint  \\
$\rm T_6$ ($\rm K$)\tablenotemark{d}               & $1430\pm290$ \\
$\rm T_7$ ($\rm K$)                                           & $1850\pm300$  \\
$\rm T_8$ ($\rm K$)                                           & $1920\pm210$  \\
$\rm \chi^{2}_{red}$\tablenotemark{e}                & $1.25$                \\
$\rm BIC$\tablenotemark{f}                                 & $66.1$                \\
\enddata
\tablenotetext{a}{The parameter symbols are defined in Table~\ref{tab:param}.} 
\tablenotetext{b}{Parameter values based on the MCMC run with no clouds and only the four main molecules: CH$_4$, CO, CO$_2$ and H$_2$O.} 
\tablenotetext{c}{95\% upper limits based on the MCMC histograms for these parameters.}
\tablenotetext{d}{Temperature parameters T$_1$ through T$_5$ and T$_9$ are unconstrained. We define the temperature parameter 
to be ``constrained'' if the $1\,\sigma$ uncertainties from the MCMC run are less than 500\,K.}
\tablenotetext{e}{The reduced $\rm \chi^{2}$ value of the best fit model from the MCMC run.} 
\tablenotetext{f}{The Bayesian Information Criterion value of the best fit model from this MCMC run.} 
\label{tab:results}
\end{deluxetable}

\begin{figure}
  \centering
  \includegraphics[scale=0.4]{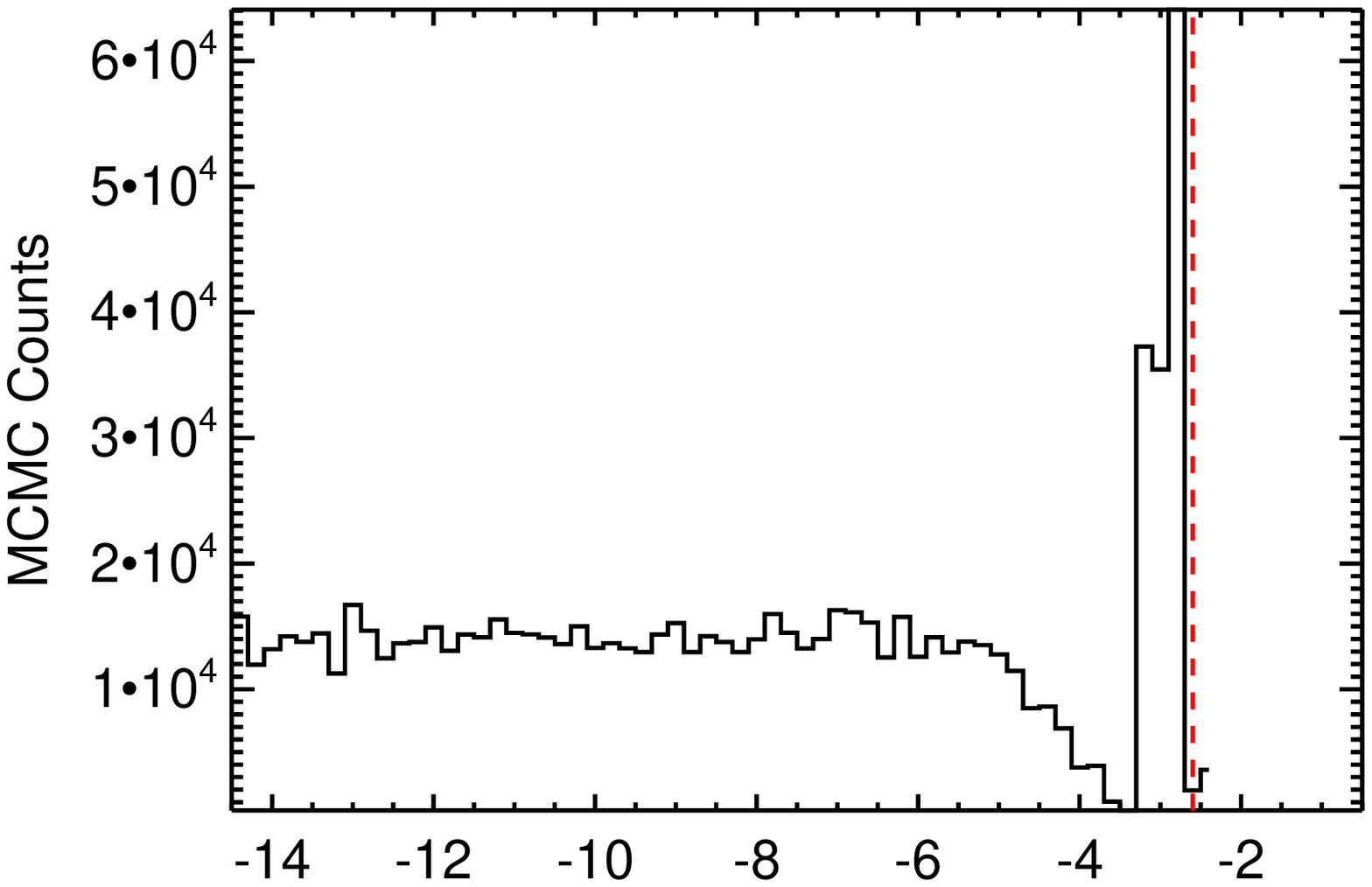}
  \caption{Similar to Figure~\ref{fig:h2ohist} but for the $\rm CH_{4}$ volume mixing ratio for the cloud-free, four molecules case (case 1). 
    Despite the peak near $\rm log(n_{CH_4}) \approx -2.5$, the long tail towards very low abundances is pronounced, making it difficult to 
    constrain the abundance of methane precisely. The red dashed line corresponds to the 95\% upper limit on the number fraction.
  }
  \label{fig:ch4hist}
\end{figure}

\subsection{Plausibility of Grey Clouds}
\label{sec:greyplausib}
Our main result, the volume mixing ratio of water is consistent between clear and grey cloud models, 
with and without the inclusion of additional molecules.
Within our framework, grey clouds are not required to explain the observed spectrum, but cannot be excluded in principle.
In order to check the physical plausibility of grey clouds consisting of large particles, we use typical values to estimate the thermal velocity of 
the gas in the atmosphere and the ``eddy mixing strength'' \citep[or $\rm K_{zz}$; e.g.,][]{hen13} to be $\rm K_{zz} > 10^{9}-10^{10}$. 
This value, compared to $K_{zz} \sim 10^4$ that \citet{mar12} find for HR 8799b, is likely unphysical, suggesting that the large-particle, 
grey-cloud assumption is too simplistic. However, there are proposed cloud compositions for which the grey assumption holds 
approximately over the relatively narrow wavelength range covered by our data, e.g., $\rm Na_{2}S$ \citep{mor12} and MgSiO$_{3}$, 
\citep{lee13} even for smaller particle sizes, comparable to the radiation wavelength.

\subsection{Comparison with Other Planets}
\label{sec:imp}

\begin{figure}[t!]
  \centering
  \includegraphics[scale=0.33]{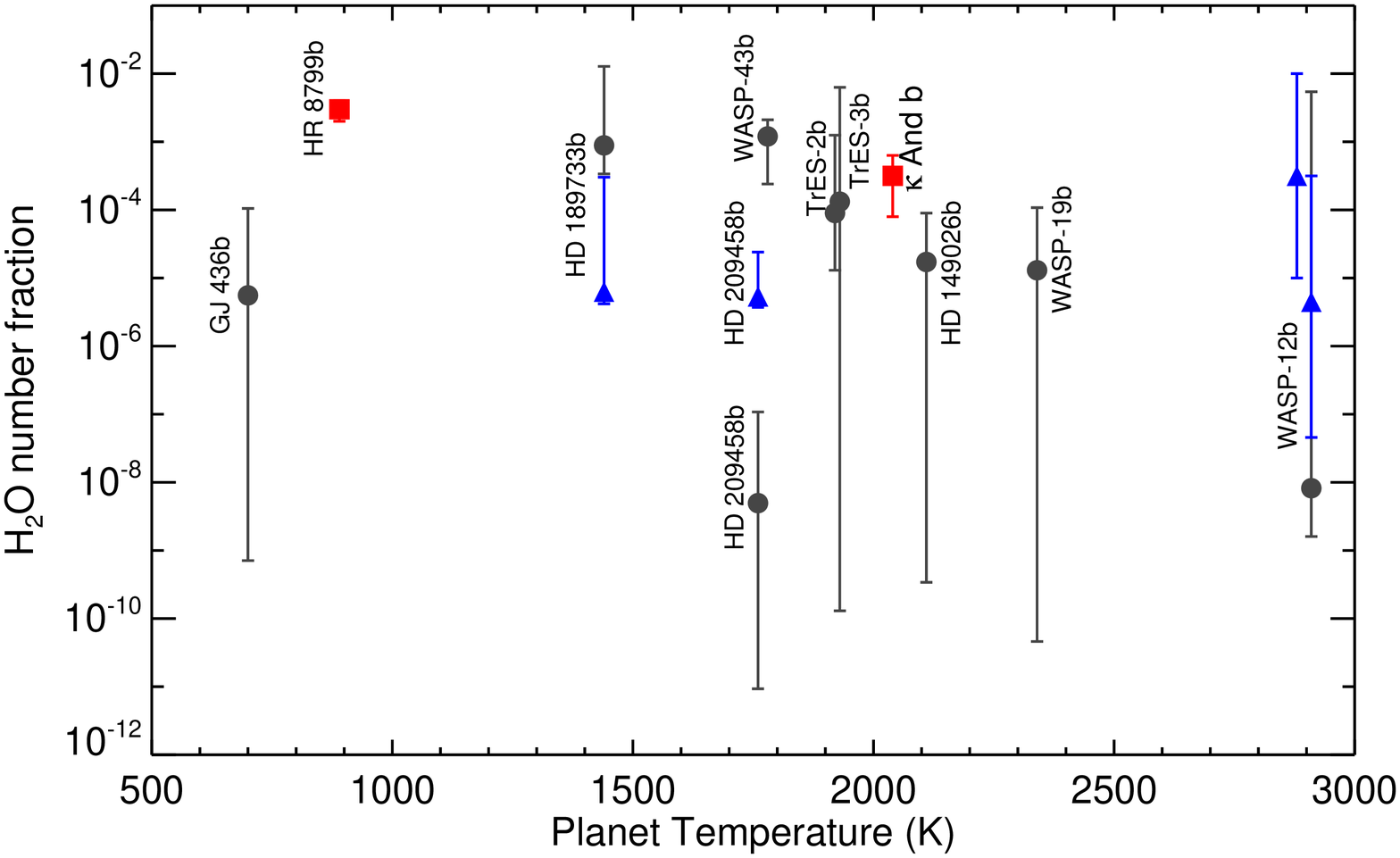}
  \caption{We compare the retrieved water abundances in the atmospheres of several hot giant planets
    with their temperatures. 
    The retrieved water abundances of hot Jupiters based on secondary 
    eclipse photometry and emission spectroscopy by \citet{lin14} are shown as black circles \citep[for WASP-43b we use
    the updated value from][]{kre14}. 
    The blue triangles without a label represent the water abundances of three planets from the \citet{lin14} sample
    based on transit spectroscopy retrieval analysis by \citet{mad14b} and \citet{kre15}. The two points 
    representing directly imaged companions \citep[this work and][]{lee14b} are shown as red squares. 
    The value we show for \kA~is based on the cloud-free models that only include the 
    abundances of CH$_4$, CO, CO$_2$ and H$_2$O as free parameters, other than the P-T profile and the surface gravity (case 1). 
    \citet{lin14} and \citet{mad14b} also make the no-clouds assumption, unlike \citet{kre14} and \citet{lee14b}.
    For this plot, we use the equilibrium dayside temperatures for the hot Jupiters, assuming no redistribution to 
    their night sides, and the effective temperatures for the directly imaged objects \citep{hin13,lee14b}.
    The uncertainties of the abundance of water in the hot Jupiters based on secondary eclipse measurements
    are typically large since most of them have only dayside photometry in the mid-infrared and no 
    spectroscopy, while the retrievals by \citet{mad14b} and those for HR~8799b and 
    \kA~relied on near-infrared spectra. This suggests that the way to maximize precision in 
    water abundance measurements is to rely on spectra of directly imaged companions. 
    The coolest ``hot Jupiter'' on the plot, GJ~436b, is in fact a 
    Neptune-sized planet orbiting a red dwarf on a $2.6$\,day orbit. The uncertainties on the water abundance of
    HR~8799b are too small to be clearly displayed in this logarithmic scale. 
  }
  \label{fig:watervsT}
\end{figure}

\begin{figure}
  \centering
  \includegraphics[scale=0.33]{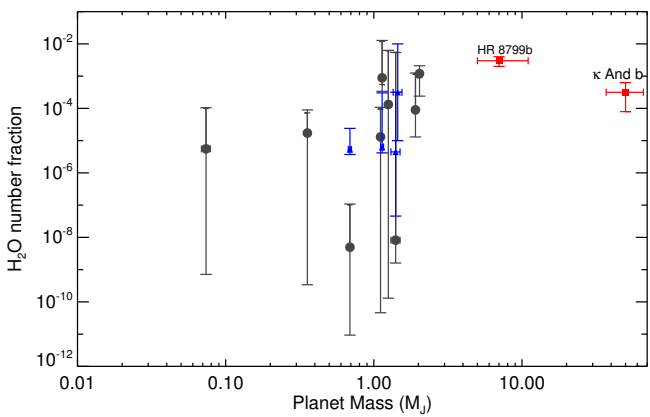}
  \caption{Similar to Figure~\ref{fig:watervsT}, but here we compare the water volume mixing ratio to the mass of the planets and substellar companions. 
  We adopt the masses listed in the The Extrasolar Planets Encyclopaedia\footnote{\url{http://exoplanet.eu/}}, except for \kA, for which we use the mass
  of $\rm 50^{+16}_{-13}\,M_{J}$ found by \citet{hin13}. Again, we see no obvious dependence of water abundance on the mass of the planet. 
  }
  \label{fig:watervsMp}
\end{figure}

We compare the water abundance we retrieve for \kA~ based on case 1 
with retrieved water abundances from other studies in Figures~\ref{fig:watervsT} and \ref{fig:watervsMp}. 
Most of the points presented in these plots assume clear cloud-less models. Despite that, there are possible 
heterogeneities in the water volume mixing ratio determination due to different assumptions, as well as due to the different methods used to obtain spectral information. 
There are also possible heterogeneities related to the temperature determination. All hot Jupiters' temperatures given in Figure~\ref{fig:watervsT} 
are day side equilibrium temperatures based on distance to the star and stellar luminosity and assuming no redistribution to the night side of the planet. 
For the two directly imaged objects, HR~8799b and \kA~we adopt the effective temperatures found by \citet{hin13} and \citet{lee14b}.

The water volume mixing ratio in the atmosphere of \kA~ is similar to the retrieved values 
from the day side emission of the transiting hot Jupiters TrES-2b ($\rm 1.3\,M_J$), TrES-3b ($\rm 1.9\,M_J$)
and HD~149026b \citep[$\rm 0.36\,M_J$,][]{sou10}, as Figures~\ref{fig:watervsT} and \ref{fig:watervsMp} suggest \citep{lin14}. 
These three hot Jupiters have day side equilibrium temperatures of $\sim1930, 1920$ and $2110$\,K, respectively, which is similar to the effective temperature
of $2040\pm60$\,K for \kA~\citep{hin13}. This is despite the fact that \kA~is likely more massive than all three of these,
and in fact than any of the other objects presented in Figures~\ref{fig:watervsT} and \ref{fig:watervsMp}. 
The water volume mixing ratio of \kA~falls within the 
range of hot Jupiter abundances, although the uncertainties in the water abundances of the transiting planets are far larger than those for 
the directly imaged planets. We attribute this to the fact that in most cases the model retrieval on secondary eclipse data relied 
on multiband photometry of the planetary day side rather than spectroscopy. In addition, 
directly imaged spectra do not suffer from systematic effects resulting 
from eclipse and transit depth measurements. The wide range of water abundances 
covered by the companions in Figures~\ref{fig:watervsT}  and \ref{fig:watervsMp} provides
no proof that the amount of water in a companion's atmosphere is correlated with its temperature, irradiation level or mass. However,
this may change as data with higher signal-to-noise ratio and with better spectral resolution and coverage become 
available for the objects, allowing for more precise abundance measurements. 

\section{Summary and Future Work}
\label{sec:con}

We find that the water abundance we derive for \kA~is not qualitatively different
from these properties derived from the day side spectra and photometry of transiting hot Jupiters, although
there is a wide range of water abundances that are currently consistent with the hot Jupiter measurements. 
Since H$_2$O is essential in the formation of the C/O ratio of an atmosphere, the similar water abundances 
may point to a common formation mechanism and location within the protoplanetary disk, followed by migration 
\citep[e.g.][]{obe11, mad14a}. This suggestion can be tested as the quality of 
secondary eclipse and direct imaging observations improves. Therefore, studying the molecular abundances 
of gas giant planets and companion brown dwarfs with very different effective temperatures, 
irradiation levels and masses that currently appear to inhabit a similar region of parameter space, 
is a logical next step. 
 
Future development of our computational approach will include proper treatment and weighting of photometric data points in addition to spectroscopy, 
which will allow full usage of all available data for a given substellar companion. While our spectrum contains 28
data points, which is a rich data set compared to many other exoplanet studies, it is still not a high-resolution, 
high-signal-to-noise spectrum. Obtaining higher quality spectra of directly imaged 
companions using the new generation of high-contrast instruments
like GPI \citep[Gemini South,][]{mac14} and SPHERE \citep[VLT,][]{beu08} would allow for better constraints
of their P-T profiles in a wider range of pressures and allow for better quantitative comparisons to the 
P-T profiles of hot Jupiters.

While the atmospheric emission model presented here is simplified and could yield 
unphysical results by not taking into account potentially 
important chemical species or physical processes, it is important to consider {\it along} with sophisticated, 
but computationally expensive, forward models. 
Our relatively fast code can account for and place uncertainties on the 
most important constituents of an atmosphere that shape the observed emission spectrum. 
The results from this and similar studies could be used as input
for forward models that include additional physical and chemical considerations. 
\\

\acknowledgements{
K.O.T., M.R.M. and S.P.Q. gratefully acknowledge the support of the Swiss 
National Centers of Competence in Research (NCCR) PlanetS initiative, 
funded by the Swiss National Science Foundation.
This research has made use of NASA's Astrophysics Data System.
This research has made use of the Exoplanet 
Orbit Database and the Exoplanet Data Explorer at \url{exoplanets.org}.
This research has benefitted from the SpeX Prism Spectral Libraries, maintained by 
Adam Burgasser at \url{http://pono.ucsd.edu/~adam/browndwarfs/spexprism}.}

\appendix
\section{MCMC Results for All Cases}

\begin{figure*}[!]
  \centering
  \includegraphics[scale=0.4]{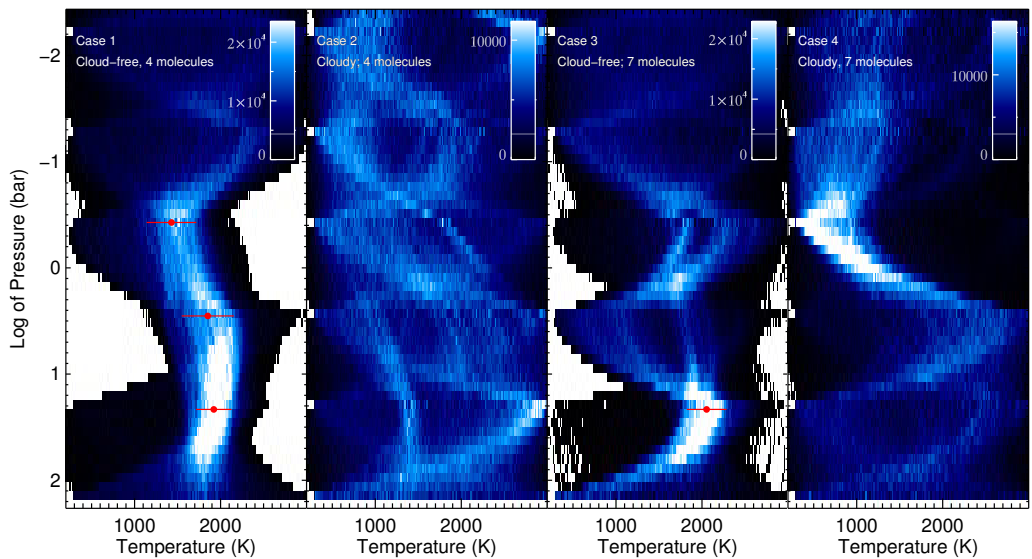}
  \caption{ Histograms of the temperature values at each pressure level from the MCMCs
    for all four considered cases: 1) without clouds and only including the four main 
    molecules: CH$_4$, CO, CO$_2$ and H$_2$O; 2) cloudy and only including the four main 
    molecules; 3) cloud-free, including the four main molecules plus the three additional ones: 
    C$_{2}$H$_{2}$, HCN, NH$_{3}$; and 4) cloudy and all seven considered molecules. 
    The red points with error bars represent the best-fit temperatures at pressures where  
    corresponding to the $\rm T_{1}-T_{9}$ free parameters. 	    
    The uncertainties are based on the corresponding MCMC histograms. We consider 
    temperature parameters with uncertainties larger than 500\,K be unconstrained and we do not plot these. 
  }
  \label{fig:bestfitPTall}
\end{figure*}

\begin{figure}[!]
  \centering
  \includegraphics[scale=0.4]{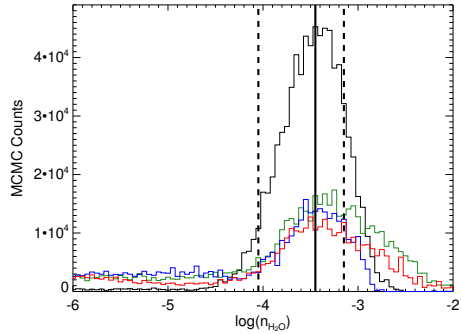}
  \caption{The histograms of cloud-free MCMC parameter parameter states for the molecular number fraction of water in log space
    for each of the four cases we consider:  cloud-free and only including the four main molecules(case 1; black); 
    cloudy and only including the four main molecules (case 2, green); cloud-free, including all seven molecules 
    (case 3, blue); and cloudy and including all seven molecules (case 4, red). All MCMC histograms peak near $\rm log(n_{H_2O}) \approx -3.5$, 
    but the cloudy or seven-molecule models have long tails towards very low abundances. We denote the
    peak of the distribution for case 1 (solid black line) and the region that covers 34\% (``1$\,\sigma$'') of the values on either
    side (dashed black lines). 
  }
  \label{fig:h2ohistall}
\end{figure}

\begin{figure*}[h!]
  \centering
  \includegraphics[scale=0.4]{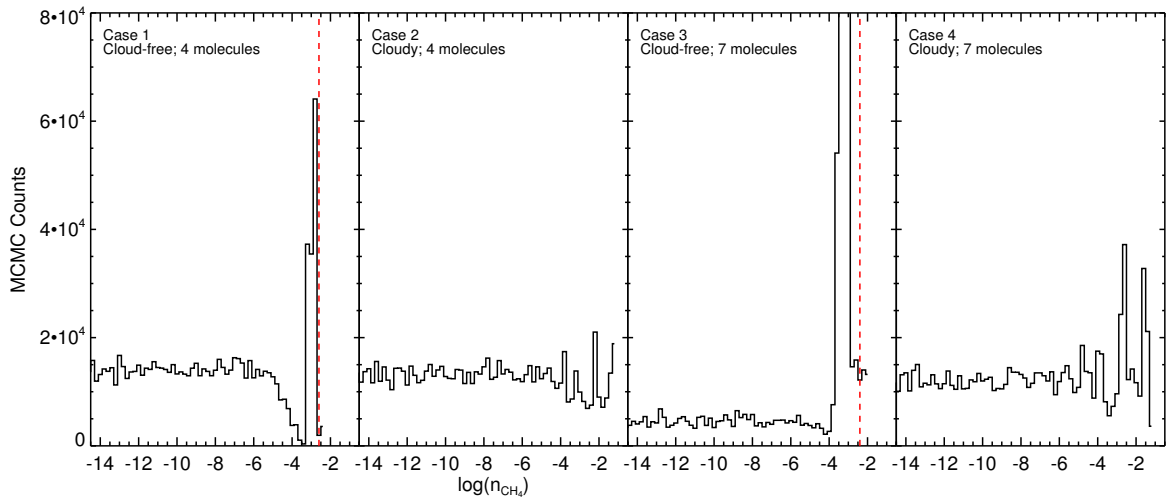}
  \caption{ Similar to Figure~\ref{fig:h2ohistall} but for the $\rm CH_{4}$ volume mixing ratios. The four panels show the MCMC 
    histograms for the four model cases labeled. In the cloudy cases the methane abundance is unconstrained. 
    In the cloud-free cases, have peaks near $\rm log(n_{CH_4}) \approx -2.5$, but their long tails towards very low abundances
    are pronounced. The red dashed lines correspond to the 95\% upper limits on the number fraction of CH$_4$ in the cloud-free models.
  }
  \label{fig:ch4histall}
\end{figure*}

Here we present the results from the MCMC runs for all four cases that we explore. Table~\ref{tab:resultsall} lists the best fit values 
for all converged physical parameters. Figure~\ref{fig:bestfitPTall} is similar to Figure~\ref{fig:bestfitPT} but shows the temperatures for 
all cases, not just for case 1. Figures~\ref{fig:h2ohistall} and \ref{fig:ch4histall} show the volume mixing ratio histograms for water and 
methane for each of the four cases. Even if we had adopted case 3 or case 4 that find upper limits on the 
number densities of C$_{2}$H$_{2}$, HCN and NH$_{3}$ these would not have been meaningful, since these molecules are not expected
to be present in such large quantities in the atmosphere of \kA. The fact that the MCMC fits place an upper limit on these values
simply confirms that these molecules play little role in the formation of the emission spectrum at this resolution.

   \begin{deluxetable*}{lcccc}[!]
   \tabletypesize{\scriptsize}
\tablewidth{0pt}
\tablecaption{Results from all MCMC Runs}
  \tablehead{
    \colhead{Parameter\tablenotemark{a}} &
    \colhead{Case 1\tablenotemark{b}} & 
    \colhead{Case 2} &
    \colhead{Case 3} & 	
    \colhead{Case 4} 
    	} 
\startdata
$\rm log(n_{H_2O})$                                           & $-3.5^{+0.3}_{-0.6}$ & $-3.3^{+0.5}_{-6.8} $  &  $-3.5^{+0.4}_{-7.8} $ & $-3.5^{+0.6}_{-7.9}$  \\
$\rm log(n_{CH_4})$\tablenotemark{c}                & $<-2.6$                    & no constraint                 &  $<-2.4$                     & no constraint                     \\
$\rm log(n_{CO}) $                                              & no constraint              & no constraint                 & no constraint                 & no constraint    \\
$\rm log(n_{CO_2})$\tablenotemark{c}                & no constraint              & no constraint                 &   $<-2.2$                    &  no constraint         \\
$\rm log(n_{C_2H_{2}})$\tablenotemark{c}          & \nodata                     & \nodata                        &  $<-2.4$                     &   $<-2.6$                     \\
$\rm log(n_{HCN}) $\tablenotemark{c}                & \nodata                    & \nodata                         &  $<-3.0$                     &   $<-2.2$  \\
$\rm log(n_{NH_{3}})$\tablenotemark{c}              & \nodata                    & \nodata                         &   $<-3.0$                    &    $<-2.6$         \\
$\rm T_6$ ($\rm K$)\tablenotemark{d}               & $1430\pm290$        & no constraint                 &  no constraint                &  no constraint  \\
$\rm T_7$ ($\rm K$)                                           & $1850\pm300$        & no constraint                 &  no constraint                &  no constraint  \\
$\rm T_8$ ($\rm K$)                                           & $1920\pm210$        & no constraint                 &  $2050\pm 230$          &  no constraint   \\
$\rm \chi^{2}_{red}$\tablenotemark{e}                & $1.25$                      & $2.21$                         & $1.13$                         & $2.06$                   \\
$\rm BIC$\tablenotemark{f}                                 & $66.1$                     & $82.1$                          & $71.3$                         & $84.4$                   \\
\enddata
\tablenotetext{a}{The parameter symbols are defined in Table~\ref{tab:param}.}  
\tablenotetext{b}{Cases: 1) cloud-free and only the four main molecules; 2) cloudy and only with the four main 
    molecules; 3) cloud-free, including the four main molecules plus the three additional ones; and 4) cloudy, with all seven considered molecules.}
\tablenotetext{c}{95\% upper limits based on the MCMC histograms for these parameters.}
\tablenotetext{d}{Temperature parameters T$_1$ -- T$_5$ and T$_9$ are never constrained. We assume that a temperature parameter is ``constrained'' if the $1\,\sigma$ uncertainties from a given MCMC run are less than 500\,K.}
\tablenotetext{e}{The reduced $\rm \chi^{2}$ value of the best fit model from this MCMC run.} 
\tablenotetext{f}{The Bayesian Information Criterion value of the best fit model from this MCMC run.} 
\label{tab:resultsall}
\end{deluxetable*}

\end{document}